\documentclass{appolb}
\usepackage{graphicx}
\usepackage{subfigure}
\newcommand{\eq}{\begin{equation}}
\newcommand{\eqx}{\end{equation}}
\newcommand{\eqn}{\begin{eqnarray}}
\newcommand{\eqnx}{\end{eqnarray}}
\newcommand{\bi}{\begin{itemize}}
\newcommand{\ei}{\end{itemize}}
\newcommand{\nn}{\nonumber}
\newcommand{\ra}{\rangle}
\newcommand{\la}{\langle}

\newcommand{\tA}{\tilde{A}}

\def\runhead{}

\begin{document}
\title{\bf Wilson loops with arbitrary charges
}


\author{Piotr Korcyl$^a$, Mateusz Koren$^{ab}$, Jacek Wosiek$^a$
\address{$^a$ M. Smoluchowski Institute of Physics, Jagellonian University\\
\L ojasiewicza St. 11,  30-348 Krakow}
\address{$^b$ Instituto de F\'{\i}sica Te\'orica UAM-CSIC, E-28049--Madrid, Spain}
}

\maketitle

\begin{flushright}
IFT-UAM/CSIC-14-120
\end{flushright}

\begin{abstract}
We discuss how to implement, in lattice gauge theories, external charges which are not commensurate with an elementary gauge coupling.

It is shown that an arbitrary, real power of a standard Wilson loop (or Polyakov line) can be defined and consistently computed in lattice formulation of
non-abelian, two dimensional gauge theories. However, such an observable can excite quantum states with integer fluxes only. Since the non-integer fluxes are not in the spectrum
of the theory they cannot be created, no matter which observable is chosen. Also the continuum limit of above averages does not exist unless the powers in question
are in fact integer. On the other hand, a new continuum limit exists, which is rather intuitive, and where above observables make perfect sense and lead to the string tension
proportional to the square of arbitrary (non necessary commensurate with gauge coupling) charge.
\end{abstract}

\section{Introduction}
Confinement remains a challenge in spite of the spectacular progress in studying non-perturbative QCD, e.g. with lattice methods.  Many intuitive models have been proposed over years to elucidate this phenomenon. In particular
the Schwinger model has been a source of valuable inspiration, also in this case. In 1975 Coleman, Jackiw and Susskind
have shown \cite{CJS} that the energy of two external charges, separated by a distance $L$, grows linearly with L for small mass, $m$, of dynamical fermions
\eq
E(L)= m e \left(1-\cos{\left(2\pi\frac{q}{e}\right)}\right) L , \label{sig}
\eqx
where $e$ is a charge of dynamical fermions and $q$ that of external sources. This result was then generalised to non-abelian systems in the large N (colour) limit with essentially the same string tension \cite{GT}, \cite{FS}. It has also a simple interpretation in terms of screening.

This paper originated in an attempt to confront Eq.\ref{sig} with lattice calculation and eventually extend it to larger masses
of dynamical fermions. However to do so one has to introduce on a lattice an external source with an arbitrary
charge, not commensurate with the elementary charge $e$ of dynamical fermion say, a quark. Surprisingly we have not found
any studies of this issue in the literature. Therefore in this paper we would like to explore even simpler question:
how to represent arbitrary real charges in a lattice version of two dimensional pure U(1) gauge theory (Quantum Maxwell Dynamics, $QMD_2$).
\section{Basics}
Partition function of $QMD_2$ on a $N_x \times N_t$ periodic lattice is known analytically
\eqn
Z&=&\int d(links)  e^{S(plaquettes)}=\Sigma_n I_n(\beta)^{N_x N_t}, \label{Z}\\
\eqnx
In the continuum limit $N_x a=L$,$N_t a = T$,$ 1/\beta=e^2 a^2 $
\eqn
Z\;\; \rightarrow\;\; \#\;\; \Sigma_n e^{-E_n T},\;\;\;\; E_n=\frac{1}{2} e^2 n^2 L,  \label{EPU}
\eqnx
Hence it is saturated by well known states of topological fluxes \cite{M}. Since there are no dynamical degrees of freedom in
two dimensions, the theory would have been trivial if not for a non trivial topology. Only integer multiples of elementary charge are allowed.

External sources are introduced by means of Wilson loops
\eqn
Z \la W(\Gamma) \ra = \int d(links)  W(\Gamma) e^{S} = \Sigma_n I_n(\beta)^{N_t N_x-n_t n_x} I_{n+1}(\beta)^{n_t n_x},  \label{W}
\eqnx
whose continuum limit
\eqn
 \Sigma_{n} \exp { \left( -\frac{e^2}{2} n^2 L (T - t ) \right)  }
 \exp { \left( -\frac{e^2}{2} \left( n^2 (L-R)+(n+1)^2 R  \right) t \right)  } \label{WC}
\eqnx
has a simple and appealing interpretation in terms of additional fluxes extending between external charges. One can easily
introduce higher charges of external sources (just replace $W[\Gamma]\rightarrow (W[\Gamma])^m$), however again, they would have to be integer multiples of $e$ and would result in replacing $n+1$ by $n+m$ in (\ref{WC}).

The obvious candidate for a pair of sources with arbitrary charge $q$ is then an arbitrary real power $(W[\Gamma])^Q$, with $Q=q/e$.
At first sight it may rise some questions of gauge invariance and non uniqueness. However a little more careful examination
shows that in fact there is nothing wrong with this proposal. Its average value can be readily calculated as above yelding

\eqn
 Z  \la W_Q \ra =  \Sigma_{m,n} I_n^{N_x N_t-n_x n_t} I_{m}^{n_x n_t}S(Q-(n-m))^{n_t+n_x}
 \label{WQ}\nn
\eqnx
with the overlap function
\eq
S(Q- n+m)=\left(\sin{\pi (Q-n+m)}/\pi (Q-n+m) \right)^2,
\eqx
resulting from the integration over two opposite links in the contour $\Gamma$.
For integer $Q$ (\ref{WQ}) reduces to (\ref{W}) with already mentioned interpretation. However for arbitrary, real $Q$ things are different as discussed in the next chapter.
\section{Q-loops}
Naively, one might expect that a Q-loop would excite a flux with an arbitrary charge $q=Qe$ with the string tension
$\sigma_Q \sim q^2$. This however is not the case. Eqs.(\ref{WQ},\ref{W}) tell us that that the {\em only} quantum states in the
system are fluxes with multiples of the elementary charge $e$. Therefore a Q-loop can excite only these states and the
value of $Q$ controls only the amplitude with which given flux is excited.

Not surprisingly this is confirmed by lattice calculations. In Fig.1 the string tension, as extracted from lattice calculation of
 $\la W_Q \ra$, is shown for a range of $Q$'s.

\begin{figure}[h]
\begin{center}
\subfigure[Monte Carlo data for charged Wilson loops.]
{\includegraphics[width=6cm]{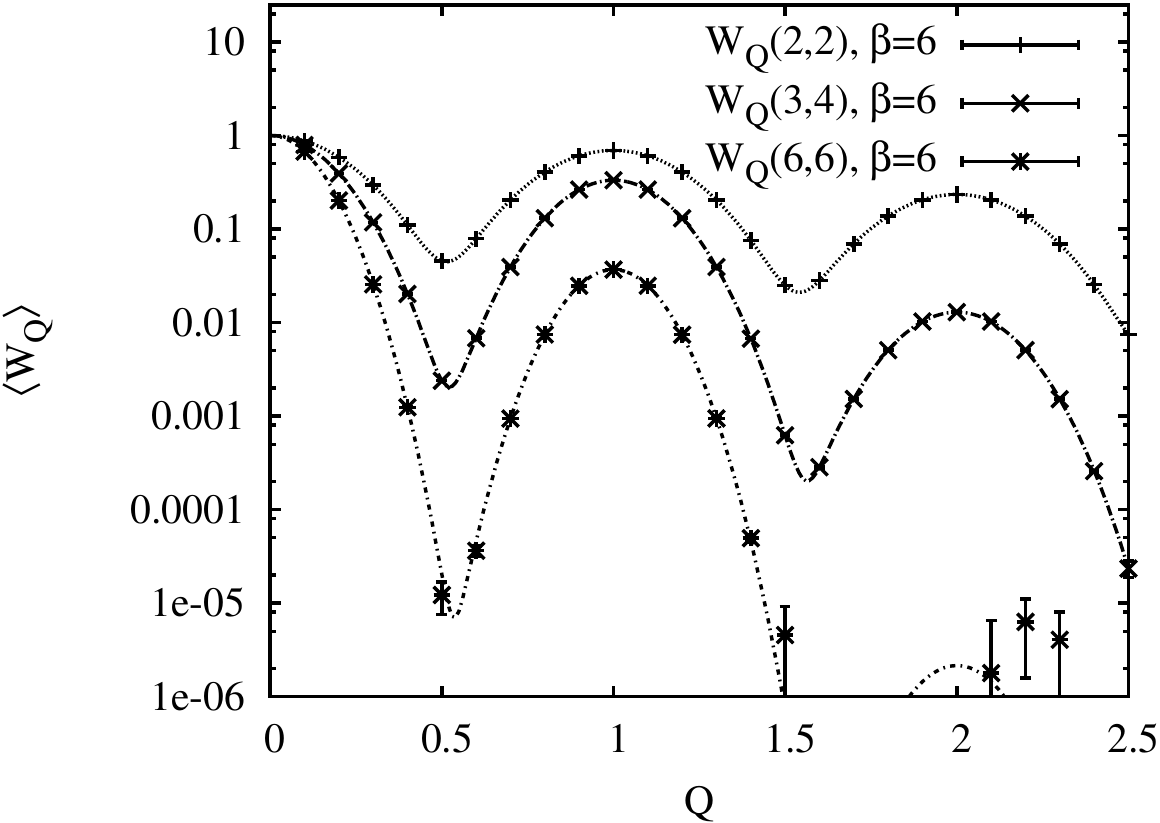}}
\subfigure[Extracted string tension $\sigma_Q$ as a function of charge for $\beta=3$ and $\beta=6$.]
{\includegraphics[width=6cm]{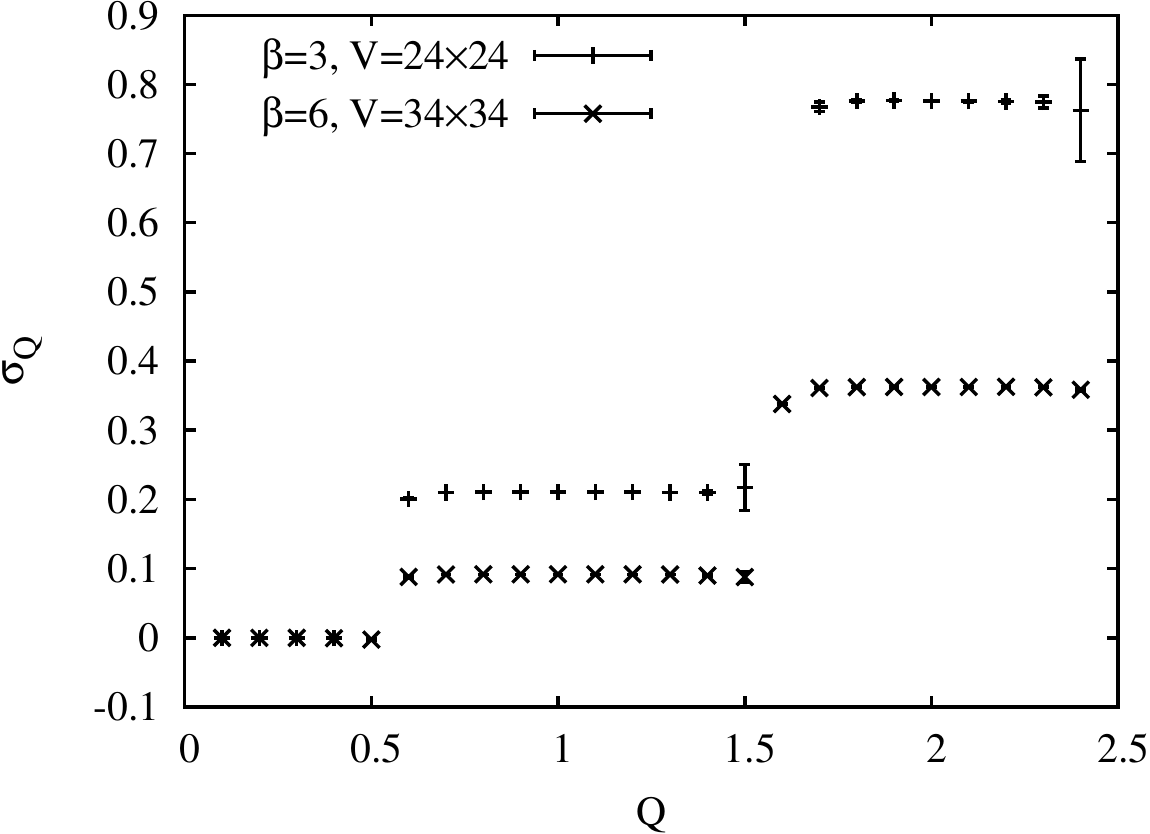}}
\caption{
Numerical data for charged Wilson loops. The HMC algorithm was employed for the simulations and lattices of
size $24 \times 24$ at $\beta = 3$ (48 million measurements) and $34 \times 34$ at $\beta = 6$ (19.2
million measurements) were used.}
\end{center}
\label{h1}
\end{figure}

For given real $Q$ the overlap $S(Q)$ is strongly peaked around the nearest integer, hence only
this flux is excited in practice. This explains the steps seen in Fig.1.

All above was done in the dimensionless  lattice units. On the other hand an attempt to obtain the continuum limit of (\ref{TMQQ}) gives
\eqn
 \Sigma_{m,n} \exp { \left( -\frac{e^2}{2} n^2 L (T - t ) \right)  }
 \exp { \left( -\frac{e^2}{2} \left( n^2 (L-R)+m^2 R  \right) t \right)  }\nn \\ \nn\\ \nn
S(Q-(n-m))^{(t+R)/a}, \nn
\eqnx
which does not exist for arbitrary $Q$.   The overlaps $S$ vanish rapidly (with $a\rightarrow0$) for arbitrary Q and are not compatible with integer-valued fluxes $m,n$ unless $Q$ is again integer. This lends credit to the original suspicion that one cannot consistently
introduce fractional (or more generally: arbitrary real) charges on a periodic lattice. However, there exists a quite natural limit where (\ref{WQ}) makes a perfect sense. It can be termed as a {\em classical limit} of large $Q$ and will be now discussed in detail.

\section{Classical continuum limit of Q loops}
For better illustration we will calculate the Green's function for the time evolution of a gauge field in the presence of two external
charges separated by a distance $R$. As in \cite{M} the continuous system is defined on a spatial circle with circumference $L$.
Consequently there is only one degree of freedom. We chose it as a constant (in x) value of the x component of a gauge field: $A_x(t)\equiv A(t)$.

On a lattice we begin with
the matrix element of the transfer matrix for two Polyakov lines separated by $n_x$ lattice units, with charge $q=Q e$.
\eqn
 G = < \{\theta'\} | \Pi_{P^Q P^Q} | \{\theta\} > = \int d\{\vartheta\}
 e^{-i Q \vartheta_0} e^{i Q \vartheta_{n_x}} e^S) \label{PiPP}
\eqnx
Similar steps as before (character expansion and integration over vertical links) give for $G$ in the Coulomb gauge
 on a lattice
\eqn
G(\Theta',\Theta) = \Sigma_{m,n} I_n^{N_x-n_x} I_{m}^{n_x}S(Q-(n-m)) e^{i n (N_x-n_x)(\Theta-\Theta')}
e^{i m n_x (\Theta-\Theta')}. \label{TMQQ}
\eqnx
Where $\Theta$ ( $\Theta'$ ) is a common, in Coulomb gauge, value of all spacial link angles in a
lower (upper) time slice.

We are now ready to define the new limit which renders the arbitrary charge $q$ meaningful.
This is the {\em classical limit}
of large $Q$, such that the actual dimensionfull   charge $q=Q e$ is finite.
It requires the gauge coupling $e$ to tend to zero in appropriate way. Moreover, the large $Q$ and small $e$ limit is taken {\em before} the continuum limit.
The limit can be roughly viewed as consisting of two steps.
In the first part $\beta$ is taken to be large, as usual, however because $e$ is small rather than the lattice constant $a$ (c.f. \ref{EPU}). This gives
\eqn
 G &=& \Sigma_{m,n} \exp { \left( -\frac{1}{2\beta} ( n^2 (N_x-n_x)+m^2 n_x ) \right)  }
S(Q-(n-m)) \nn\\ && e^{i n (N_x-n_x) \Delta\Theta} e^{i m n_x \Delta\Theta} \nn
\eqnx
Now, two things happen: 1) at large $\beta$ and fixed lattice distances, important contributions to the sum come from large fluxes
$(m,n \sim b=\sqrt{\beta} )$ , exponentials become smooth functions of $u=n/b$ and $v=m/b$, and 2) at the same time Q becomes $\sim b$ so we can write
\eqn
Q=\frac{q}{e}=\frac{b}{g},\;\;\;\;\ g=\frac{1}{q a},\;\;\;\;\ b=\frac{1}{e a}\longrightarrow\infty,
\eqnx
and obtain
\eqn
G &= &  \beta \int d u d v  \exp { \left( -\frac{1}{2} ( u^2 (N_x-n_x)+v^2 n_x ) \right)  }
S\left(b(g^{-1}-(u-v))\right) \nn\\&&e^{i b u (N_x-n_x)(\Theta-\Theta')} e^{i b v n_x (\Theta-\Theta')} \nn
\eqnx
using
\eqn
S(b \Delta) \stackrel{b\rightarrow\infty}{\longrightarrow}  \frac{1}{b}\delta(\Delta)
\eqnx
we can do one integral to obtain
\eqn
G =  \sqrt{\beta} \int d u  \exp { \left(- \frac{1}{2} ( u^2 (N_x-n_x)+(u- g^{-1}) n_x ) \right)  }
 e^{i u (N_x-n_x)(\tA-\tA')} e^{i (u- g^{-1}) n_x (\tA-\tA')} \nn
\eqnx
where we have also rewritten the phase factors in terms of the continuum field $\tA=A/e$, $\Theta_L=e a N_x \tA$.
Now do the gaussian integral an take the continuum limit. To this end, rearrange the
 quadratic terms and the phase factors, $\rho=n_x/N_x$,
\eqn
 G =   \sqrt{\beta}\left\{ \int d u  \exp { \left( -\frac{1}{2}  (u- \rho/g)^2 N_x \right)  }
 e^{i  (u- \rho/g) N_x \Delta \tA} \right\}  \exp { \left( -\frac{1}{2}  g^{-2} \rho(1-\rho) N_x \right)  }\nn
\eqnx
which gives finally
\eqn
G(\tA',\tA,\epsilon) =  \sqrt{\beta}  \sqrt{\frac{2\pi a}{L}}
 \exp {\left( -\frac{L}{2}\frac{(\tA-\tA')^2}{a} \right) }  \exp { \left( -\frac{q^2}{2}  \rho(1-\rho)  L a \right)  }\nn
\eqnx
which is again proportional to the kernel for propagation of a free particle over an infinitesimal time lapse $\epsilon=a$,  but now the propagation takes place in a constant background potential
\eqn
V=\frac{q^2}{2}  \rho(1-\rho)  L      \label{V}
\eqnx
with arbitrary, real value  of a classical charge $q$.

Notice that the coordinate $\tA$ is now not periodic. Periodicity $\tA \rightarrow \tA + \frac{2\pi}{ L e }$
was lost while taking the $e \rightarrow 0$ limit. For the same reason, the discrete spectrum of topological fluxes
has turned into the continuous one of free momenta $ u - \rho/g $ .

However some memory of the periodic nature of the microscopic system remains. Namely the
effective string tension
\eqn
\sigma = \frac{q^2}{2}  \rho(1-\rho)
\eqnx
vanishes at $R=0,L$. At these configurations the microscopic string between external sources begins/completes
a single winding around the circle increasing the effective flux by one quantum. Since $V$ represents the {\em difference} between
the sum of the two fluxes (inside and outside the pair of sources) and the effective single flux around the circle, it vanishes at
$R=0$ and $L$. In another words: integer, and only integer, (in terms of winding) fluxes are "screened"  (or better: accommodated) by the intrinsic spectrum of periodic $QMD_2$.

Given existence of the new limit, it is also important to estimate how fast (or slow) it can be reached in practice.
This can be answered by confronting the analytical predictions with actual MC simulations. Similar
calculations for the trace of (\ref{PiPP}) give the easily testable prediction for the MC average of two Polyakov lines
on a lattice with unit length in time direction.
\eqn
\la P^{\dagger}_Q P_Q \ra  \stackrel{\beta\rightarrow\infty}{\longrightarrow}  \exp { \left( -\frac{1}{2}  q^{2} a^2
\rho(1-\rho) N_x \right)}
\eqnx
The limit is taken with  $Q^2=q^2 a^2 \beta$ with $q a$ kept fixed.
In the continuum above prediction reads
\eqn
\la P^{\dagger}_Q P_Q \ra \stackrel{cont.\;lim}{\longrightarrow} \exp { \left( -\frac{1}{2}  q^{2} \rho(1-\rho) L a \right)}.
\label{WQcont}
\eqnx
The above can be generalized for an arbitrary $N_t \times N_x$ lattice and we obtain
\eqn
\la P^{\dagger}_Q P_Q \ra  \stackrel{\beta\rightarrow\infty}{\longrightarrow}  \exp { \left( -\frac{1}{2}  q^{2} a^2
\rho(1-\rho) N_x N_t \right)}
\eqnx
and
\eqn
\la P^{\dagger}_Q P_Q \ra \stackrel{cont.\;lim}{\longrightarrow} \exp { \left( -\frac{1}{2}  q^{2} \rho(1-\rho) L T \right)}.
\label{WQcontrect}
\eqnx

\begin{figure}[h]
\begin{center}
\includegraphics[width=12cm]{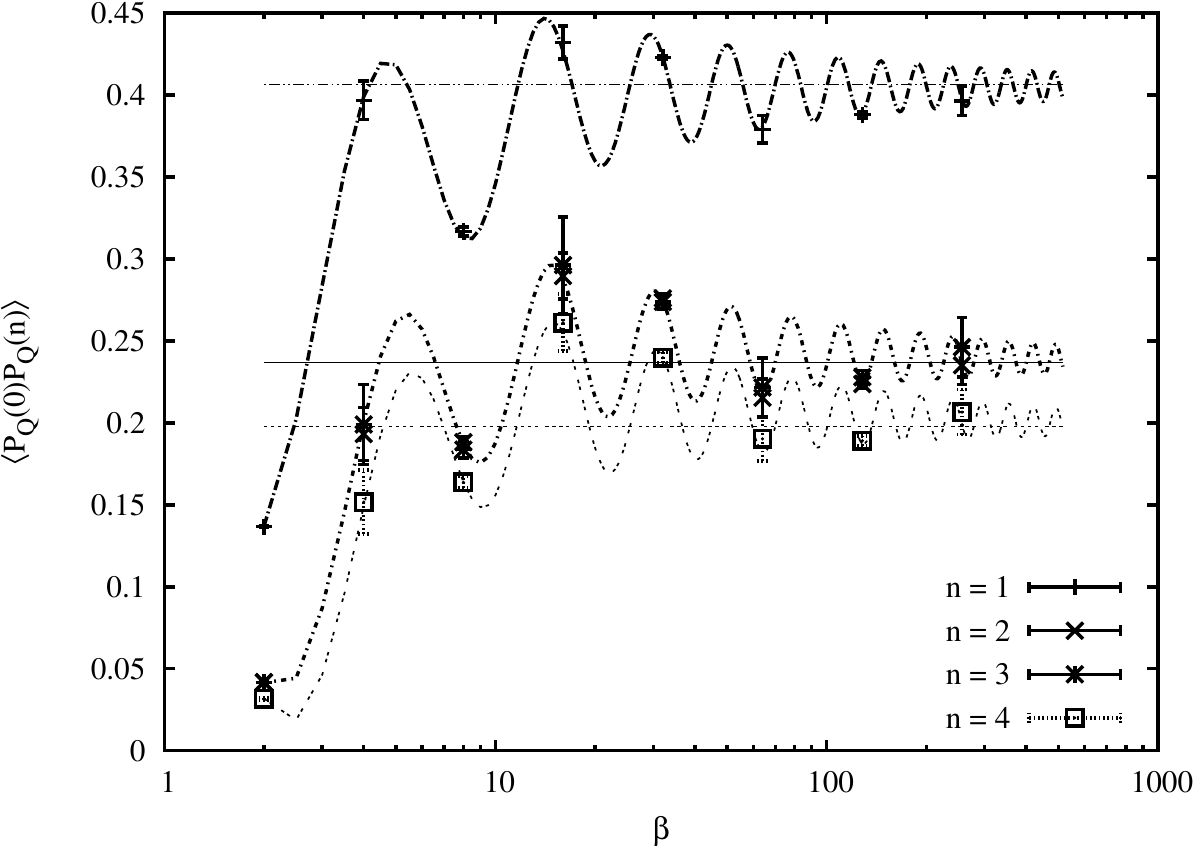}
\end{center}
\vskip-4mm \caption{Comparison of $\la P^{\dagger}_Q P_Q \ra$ measured on a $6 \times 6$ lattice (data points) with the
analytic prediction of Eq.\ref{WQ} (solid curves) and with the continuum limit given by Eq.\ref{WQcontrect} (horizontal lines).}
\label{h2}
\end{figure}

This is tested in Fig. \ref{h2} for $Q=0.85/\sqrt{2}$ by a simulation on a $6 \times 6$ lattice. The actual
numerical implementation profited from the possible equivalence with a spin chain of length 36 \cite{sinclair}. The latter was
simulated using the cluster algorithm \cite{wolff} which practically eliminates the critical slowing-down problem and the errors
were estimated using the $\Gamma$-method following Ref.\cite{errors}. Simulated points perfectly agree with Eq. \ref{WQ} which
depicted in Fig. \ref{h2} as a solid curve ($n_t$ was set to $N_t$ since we are dealing with two Polyakov loops separated by a
distance $n_x$). The positions of the horizontal lines were calculated using Eq.\ref{WQcontrect} and correspond to the continuum
prediction.

\section{Conclusions}
Arbitrary, real powers of Wilson loops (or Polyakov lines) are natural candidates if one wishes to study, on a lattice, external charges which are not commensurate with the elementary gauge coupling. However introduction of such observables raises some subtle questions which show up even in the simplest gauge models. We have discussed them in the case of
 Quantum Maxwell Dynamics - a pure gauge U(1) theory in two dimensions.

 Q-loops, as we call them, can be consistently defined on a lattice: results of MC simulations fully agree with the analytical predictions which are readily available in this simple model.

 They do not excite/create quantum states (fluxes) with arbitrary real charge, however. Such states do not exist in the theory. Instead a Q-loop excites mostly the quantum state of flux with the integer charge which is closest to the "charge carried by a Q-loop". This charge controls the amplitude with which integer (in units of elementary charge) charges are excited.

 Above applies to the lattice. The continuum limit of Q-loops does not exist unless a charge of Q-loop is multiple of
 the elementary gauge coupling.

 However a modification of the continuum limit is possible where Q-loops are well defined and give rise to some interesting
 physics. The modification is inspired by classical considerations and applies when the arbitrary  charge of external sources
 is much larger than that of the elementary flux. Then the discrete quantum spectrum becomes continuous and the effective fluxes with arbitrary charge emerge. Existence of this effective, classical behaviour was also confirmed numerically and the ranges of parameters where it is seen were estimated.

{\em Acknowledgements}
This work is supported partially through NCN grant nr 2011/03/D/ST2/01932 and by the Foundation's for Polish Science MPD Programme
co-financed by the European Regional Development Fund, agreement no. MPD/2009/6. PK acknowledges support of the Foundation's for Polish Science
fellowship START. MK acknowledges financial support from the grants FPA2012-31686 and the MINECO Centro de Excelencia Severo Ochoa Program
SEV-2012-0249, and participation in the Consolider-Ingenio 2010 CPAN (CSD2007-00042).

The numerical computations were carried out with the equipment purchased thanks to the financial support of the European Regional Development Fund in the framework of the Polish Innovation Economy Operational Program (contract no. POIG.02.01.00-12-023/08).



\end{document}